\documentclass[twocolumn,showpacs,preprintnumbers,amsmath,amssymb,prl,superscriptaddress]{revtex4-1}

\usepackage{amsfonts}
\usepackage{amsmath}
\usepackage[makeroom]{cancel}
\usepackage[english]{babel}
\usepackage[T1]{fontenc}
\usepackage{times}
\usepackage{mathrsfs}
\usepackage{graphicx}
\usepackage{dcolumn}
\usepackage{bm}
\usepackage{wasysym}
\usepackage[colorlinks,bookmarks=true,citecolor=blue,linkcolor=red,urlcolor=blue]{hyperref}
\usepackage[tight, FIGTOPCAP, hang, raggedright, nooneline]{subfigure}
\usepackage{hyperref}
\usepackage{xcolor}
\usepackage{epstopdf}
\usepackage{epsfig}
\usepackage{amssymb}

 \definecolor{Green}{RGB}{80,182,0}

\usepackage{color}

\newcommand{\la}{\langle}
\newcommand{\ra}{\rangle}

\begin{document}
\title{Neutron scattering signatures of the 3D hyper-honeycomb Kitaev quantum spin-liquid}
\author{A.~Smith}
\author{J.~Knolle}
\author{D.~L.~Kovrizhin}
\affiliation{T.C.M. Group, Cavendish Laboratory, J. J. Thomson Avenue, Cambridge CB3 OHE, United Kingdom}
\author{J.~T.~Chalker}
\affiliation{Theoretical Physics, Oxford University, 1, Keble Road, Oxford OX1 3NP, United Kingdom}
\author{R.~Moessner}
\affiliation{Max Planck Institute for the Physics of Complex Systems, D-01187 Dresden, Germany}
\date{\today}

\begin{abstract}
Motivated by recent synthesis of the hyper-honeycomb material
$\beta$-$\mathrm{Li_2 Ir O_3}$, we study the dynamical structure factor (DSF) of 
the corresponding 3D Kitaev quantum spin-liquid (QSL), whose fractionalised degrees of freedom are Majorana fermions and emergent flux-loops. 
Properties of this 3D model are known to differ in important ways from those of its 2D counterpart -- it has finite-temperature phase transition, as well as distinct features in Raman response. We show, however, that the qualitative behaviour of the DSF is broadly dimension-independent. Characteristics of the 3D DSF include a response gap  even in the gapless QSL phase and an energy dependence deriving from the Majorana fermion density of states.  
Since the majority of the response is from states containing a single Majorana excitation, our results suggest inelastic neutron scattering as the spectroscopy of choice to illuminate the physics of Majorana fermions in Kitaev QSLs.\end{abstract}

\maketitle

{{\textit{Introduction.}} A central theme in an ongoing search for new states of matter is the interplay of \textit{interactions} and \textit{dimensionality}, leading to new exotic phases displaying e.g.~topological order and quantum number fractionalisation. Prime examples are magnets in which quantum fluctuations (QF) are strong enough to suppress long-range magnetic order, giving rise instead to quantum spin-liquids. One route to these is via reduced dimensionality, as shown by many experimental examples of QSLs in spin chain systems. For such 1D systems, the materials realisation and theoretical understanding of spin-liquid behaviour is well established \cite{gogolin2004bosonization,giamarchi2003quantum}. In higher dimensions an alternative route is required, and Anderson in his original proposal \cite{AndersonRVB} of resonating valence bond (RVB) states recognised the importance of frustration. Here, however, physical examples are scarce, and theoretical understanding is less complete. Indeed, it took several decades before even the existence of RVB phases for microscopic 2D model Hamiltonians was firmly established theoretically~\cite{Moessner2001}. 

Theoretical tools and solvable models  are important for developing the understanding of unconventional phases. A number of powerful methods exist for 1D, such as  density matrix renormalisation group (DMRG) \cite{White1993} and  Bethe-Ansatz, which allow for a {\it quantitative} comparison between theory and experiment \cite{Mourigal2013435}. For 2D and 3D QSLs, a qualitative understanding has been developed using large-$N$ limits \cite{Read1989,Read1991}, appropriate mean field theories \cite{Wen1991} and the identification of long-wavelength descriptions \cite{Moessner2003,Hermele2004}. These approaches do not, however, provide detailed results for specific realistic models. 

Against this background, the Kitaev honeycomb model \cite{Kitaev} occupies an important position as a rare instance of an exactly solvable 2D model that supports a variety of different QSL phases, stabilised by exchange frustration. Moreover, its 3D generalisations \cite{MandalKitaev} provide an opportunity to examine the influence of dimensionality on these phases.

Materials that are candidates for the realisation of Kitaev frustration should have dominant spin-orbit coupling, since exchange interactions in the model link together real-space and spin-space anisotropies. 
Recently, a set of 2D honeycomb lattice compounds, \{Na,Li\}$_2$IrO$_3$ iridates \cite{Jackeli,Singh2010,Singh2012,Modic,Takayama} and RuCl$_3$ \cite{Plumb2014,Sears2015,Majumdar2015,Sandilands2015,Sik2015,Banerjee2015} have been proposed 
to exhibit dominant Kitaev-like spin exchange arising from spin-orbit interactions \cite{Rau2015}. 

In addition, in two insulating 3D polymorphs $\beta$- and $\gamma$-Li$_2$IrO$_3$ \cite{Takayama,Modic}, the Ir$^{4+}$ have been shown to form structures dubbed hyper-honeycomb and stripy-honeycomb lattices.  These are part of a whole series of ``harmonic" honeycomb lattices which might realise Kitaev QSL physics in 3D  via the interplay of spin orbit coupling,  interactions and the simple fact that each lattice site has coordination number three~\cite{Kimchi,LeeHarmonic,Modic,Lee,HermannsQSL,HermannsWeyl}. Recent work  \cite{Kim}
has identified $j_\text{eff} = 1/2$ degrees of freedom in the low-energy description of $\mathrm{\beta\text{-}Li_2 Ir O_3}$ %
and furthermore confirmed that the effective spin-model has dominant Kitaev-like exchange. 
So far, all experimental candidate materials show long-range magnetic order at low temperatures. However, an observation of non-coplanar spiral magnetism in 3D polymorphs is in itself a strong indication of  dominant Kitaev exchange \cite{Biffin2014,Biffin2014b,Kimchi2015}, and perhaps a proximate QSL. 
Also, it has been shown that the QSL of the Kitaev model is stable with respect to small integrability-breaking Heisenberg perturbations, while larger perturbations result in various magnetically ordered states \cite{Kimchi,Kim,LeeHarmonic}.

Here, we present 
the results of an exact calculation 
of the dynamical spin response for a 3D QSL, the Kitaev QSL of the hyper-honeycomb lattice \cite{MandalKitaev} of $\mathrm{\beta\text{-}Li_2 Ir O_3}$. Our results show signatures of spin fractionalisation into emergent quasiparticles, Majorana fermions and flux-loops.
The character of the QSL phase itself reflects the dimensionality of the lattice. In particular, point like fluxes of the honeycomb lattice in 2D are replaced by extended flux-loops in 3D, which leads to a true finite-temperature phase transition \cite{Nasu2014}. 

Our central finding is that the dynamical structure factor (DSF) is a rather direct probe of the Majorana fermion excitations in Kitaev models, independently of dimension. While details vary -- in 3D, fine structure from several Majorana bands is present, while in 2D features that stem from van Hove singularities in the Majorana density of states (DOS) are more prominent --  
the changes are small.
This is in  contrast to other dynamical probes, such as Raman scattering, which shows qualitative differences between 2D and 3D Kitaev QSLs, e.g.~via polarization dependence \cite{Knolle2014,Perreault2015}. 

The DSF measures the excitations induced in the ground state by a spin flip. In Kitaev models these consist of static fluxes and a variable number of Majorana excitations. This perspective has already enabled  us to develop a complete theoretical picture for a dynamical response in all different gapped and gapless, Abelian and non-Abelian QSL phases of the {\it 2D} honeycomb Kitaev model \cite{PRL,Knolle2015};  in fact, signatures of the Majoranas have arguably already been observed in the short time dynamics (high energy response) of RuCl$_3$ compounds \cite{Sandilands2015,Banerjee2015}. Calculations for the 3D model present a significant additional challenge, made feasible only by the technical developments we have described elsewhere \cite{Knolle2015}.

 \begin{figure}[b]
\includegraphics[width=.45\textwidth]{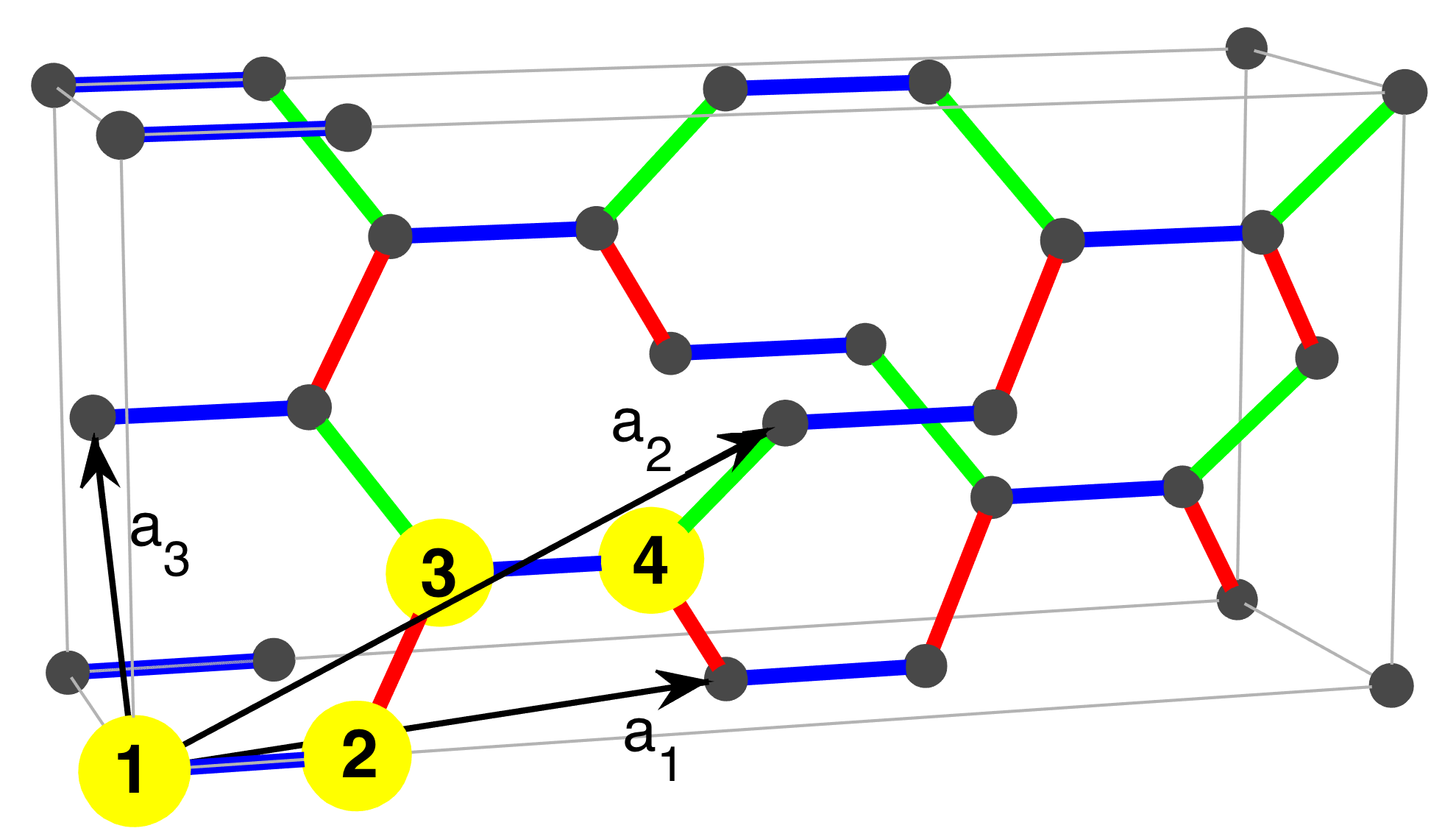}
\caption{Schematic picture of the unit cell of a hyper-honeycomb lattice showing three inequivalent bond types $x,y,z$ together with lattice vectors $\mathbf{a}_i$. Yellow sites make up the primitive unit cell consisting of four $\mathrm{Ir}$ atoms in $\beta$-$\mathrm{Li_2 Ir O_3}$.} \label{fig: lattice}
\end{figure}

\paragraph{\textit{Model.}} The hyper-honeycomb 3D Kitaev model is built from spin-$1/2$ degrees of freedom arranged on a hyper-honeycomb lattice (see Fig.~\ref{fig: lattice}) that interact via bond-dependent, nearest-neighbour Ising exchange $J_a$. Three lattice directions are labelled by $a = x,y,z$ referring to components of spins involved in the exchange bond $a$. In terms of Pauli matrices $\hat{\sigma}^a_j$ and using the notation $\la j k \ra_a$ to indicate two sites, $j$ and $k$, connected via a bond $a$, the Hamiltonian reads
\begin{equation}
\hat{H} = - \frac{J_x}{2} \sum_{\la j k \ra_x} \hat{\sigma}^x_j \hat{\sigma}^x_k - \frac{J_y}{2} \sum_{\la j k \ra_y} \hat{\sigma}^y_j \hat{\sigma}^y_k - \frac{J_z}{2} \sum_{\la j k \ra_z} \hat{\sigma}^z_j \hat{\sigma}^z_k.\label{KM}
\end{equation}
The model has two types of ground state, namely gapless and gapped QSLs for $|J_\alpha| \lessgtr |J_\beta| + |J_\gamma|$ (with $\alpha \neq \beta \neq \gamma\neq\alpha$) \cite{MandalKitaev}, and the topology of the phase diagram is the same as in a number of other 2D and 3D Kitaev models~\cite{Kitaev,HermannsQSL,Kimchi}. We note that the Majorana spectrum in 2D honeycomb/3D hyper-honeycomb possesses a single Dirac point, and a gapless nodal line respectively.

Our calculation has as its starting point the re-expression of this Hamiltonian, following the original approach of Kitaev \cite{Kitaev},
in terms of Majorana fermions. Of this we briefly mention the features important for understanding the physical content of our results;
we discuss what is technically  new in three dimensions below, and refer the reader to the methods we have developed in the context
of two dimensions Refs.~\cite{PRL,Knolle2015}. 

The spin on site $j$ is expressed in terms of four Majorana fermions: $\hat{c}_j, \hat{b}^a_j$, with $a = x,y,z$. The spin operators can then be represented as $\hat{\sigma}^a_j = i \hat{c}_j \hat{b}^a_j$. The two `species' of Majorana fermions  take on separate roles. The $\hat b$'s combine to yield
emergent non-dymanic fluxes. These are captured via loop operators $\mathcal{W}_l$; see Refs.~\cite{Lee,MandalKitaev} for their precise definition in terms of spin operators. The loop operators have eigenvalues $\mathcal{W}_l = \pm 1$ associated with the $\mathbb{Z}_2$-fluxes piercing the irreducible loops. 
These fluxes are point-like excitations in 2D, but form lines in 3D.
The $\hat{b}$ Majorana fermions are complemented by the $\hat c$'s, which can be combined in pairs to make standard complex dynamical fermions. Using the conserved quantities corresponding to loop degrees of freedom, the Hilbert space can be decomposed into a `flux', $|F\rangle$ and a `matter', $|M\rangle$, sector. For a given flux sector, 
a Hamiltonian for the matter fermions describes hopping between sites in the presence of the  $\mathbb{Z}_2$ gauge fluxes. In the following, we denote the ground state of $\hat{H}$ (which is flux-free) as $| 0 \ra = |F_0\ra \otimes |M_0\ra$.

\paragraph{\textit{Dynamic structure factor.}} Our central result is the numerically exact evaluation of the DSF $S(\mathbf{q},\omega)$, defined as
\begin{equation} \label{eq: structure factor}
S(\mathbf{q},\omega) = \frac{1}{N} \sum_{a,jk} e^{-i\mathbf{q}\mathbf{r}_{jk}}\int dt \; e^{i\omega t} S^{aa}_{jk}(t),
\end{equation} 
where $S^{ab}_{jk} = \la 0 | \hat{\sigma}^a_j(t) \hat{\sigma}^b_k(0) | 0 \ra$ is the time-dependent spin correlation function. The DSF is directly related to the cross-section measured in inelastic neutron scattering (INS) experiments \cite{Lovesey}, and at $\mathbf{q}=0$ to the signal obtained in ESR experiments. 
At the heart of our analysis is the observation that the static flux degrees of freedom (the $\hat b$'s) can be eliminated from the calculation which leads to strictly zero spin correlations beyond nearest neighbours and a local quantum quench of Majorana fermions~\cite{BaskaranExact}: a Majorana fermion $\hat{c}_k$ is added to the ground state $|M_0 \ra$ of $\hat{H}_0$ which then evolves under the action of a matter  Hamiltonian corresponding to a different flux sector. For each component $S^{aa}_{jk}$ this flux sector is obtained by flipping the $\mathbb{Z}_2$ gauge field on the corresponding a-lattice bond $\la i j\ra_a$. Note that the increased number of inequivalent bonds in the hyper-honeycomb lattice compared to the honeycomb one entails an increased number of inequivalent local quench problems entering Eq.~\ref{eq: structure factor}.
Despite the exact  solvability of the model, the non-equilibrium nature of the response makes it notoriously difficult to calculate the DSF. While in 2D one can resort to finite-size numerical calculations~\cite{Knolle2015}, these become useless for 3D. Using the exact semi-analytical
approach \cite{PRL} which we developed in the studies of the 2D case we have been able to solve the problem
also in 3D in the thermodynamic limit. 
\begin{figure}[t]
\includegraphics[width=.5\textwidth]{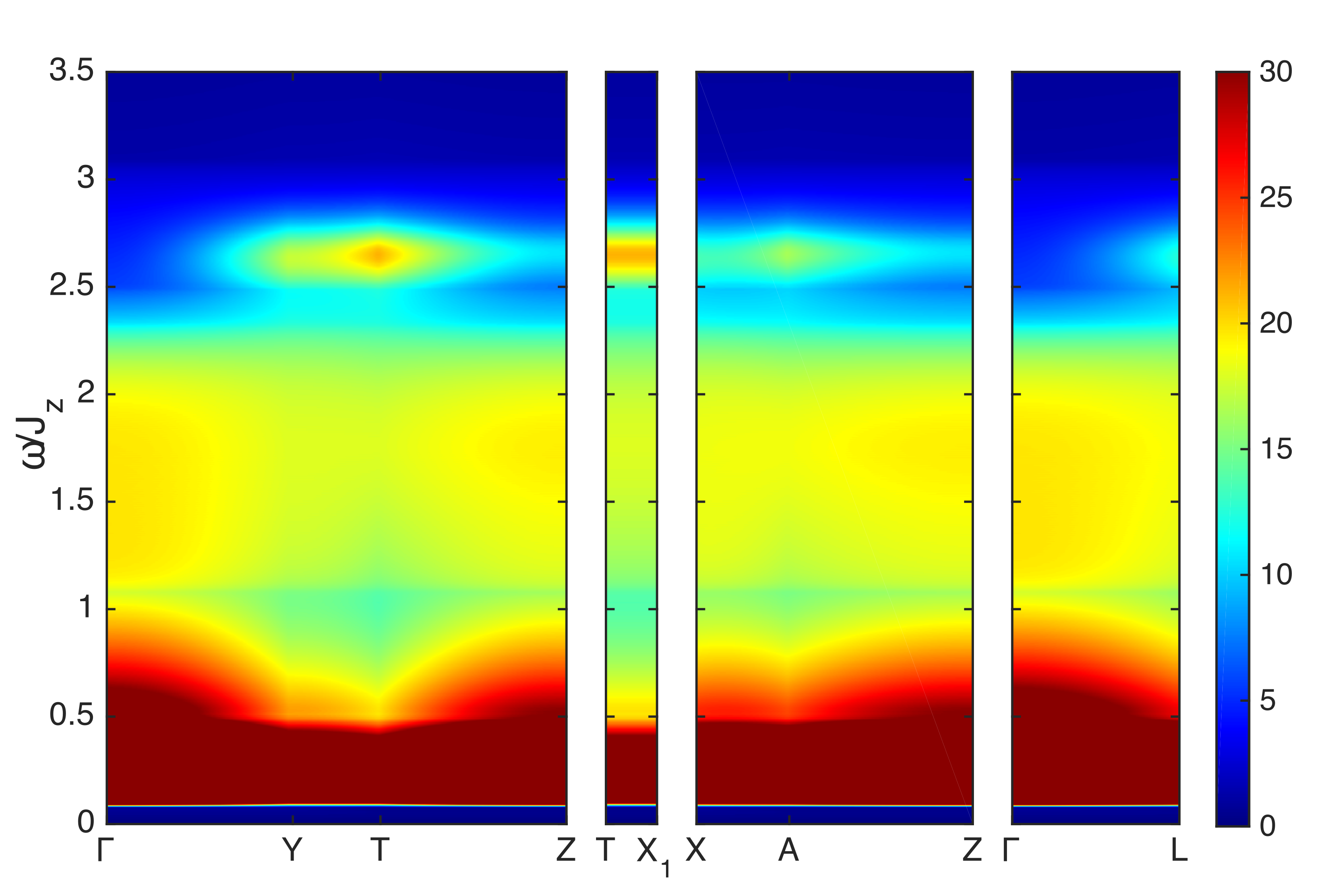}
\caption{Dynamical structure factor as a function of $\omega/J_z$ along four high symmetry cuts in the Brillouin zone, shown for the isotropic point $J_x=J_y=J_z$ on a linear colour scale. Note that in order to improve the visibility of the lower intensity features especially at high energies all intensity values above 30 (arb. units) are shown in dark red. For the full range of intensity see Fig.~\ref{fig: components}.} \label{fig: full struct}
\end{figure}

The Lehmann representation of the DSF provides a number of qualitative insights. Let $\{| \lambda \ra\}$ be the many-particle eigenstates of the matter Hamiltonian in the presence of four flux-loops with a common a-bond. Then
\begin{equation}\label{eq: Lehmann}
S^{aa}_{jk}(\omega) = 2\pi F^{a}_{jk} \sum_{\lambda} \la M_0 | \hat{c}_j | {\lambda} \ra \la {\lambda} | \hat{c}_k |M_0 \ra \delta(\omega - \Delta E_\lambda),
\end{equation}
where $F^{a}_{jk} \in \{ 1,i,-i\}$ depend on the spin component, and $\Delta E_\lambda$ is the excitation energy, which includes the contribution from the flux-gap. %
This representation reveals the main qualitative features of the response. They are remarkably independent of the dimensionality. First, the response vanishes below the flux-gap, which is the energy cost of exciting flux-loops by flipping a bond. Strikingly, this response gap is present even in the QSL phase with gapless Majorana excitations. Note that the value of the flux-gap depends significantly on the spin component probed if the $J_a$'s are not all equal.
The gap is the most direct consequence of fractionalisation, and is the signature feature of the emergent $\mathbb{Z}_2$ gauge field. 
The response above the flux-gap is entirely due to matter fermions. Although it includes multi-particle contributions, we find that the single-particle sector is dominant (accounting e.g. for 87\% of the response at ${\bf q}=0$ with all $J_a$ equal). INS therefore probes the Majorana fermion DOS rather directly.
\begin{figure}[b]
\includegraphics[width=.5\textwidth]{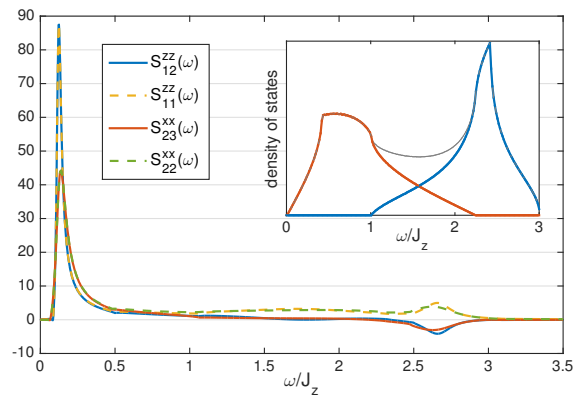}
\caption{The four inequivalent components of the dynamical structure factor at the isotropic point $J_x = J_y = J_z$. Inset: band resolved DOS of the Majorana fermion dispersions.} \label{fig: components}
\end{figure}

\begin{figure*}[t]
\includegraphics[width=1.0\textwidth]{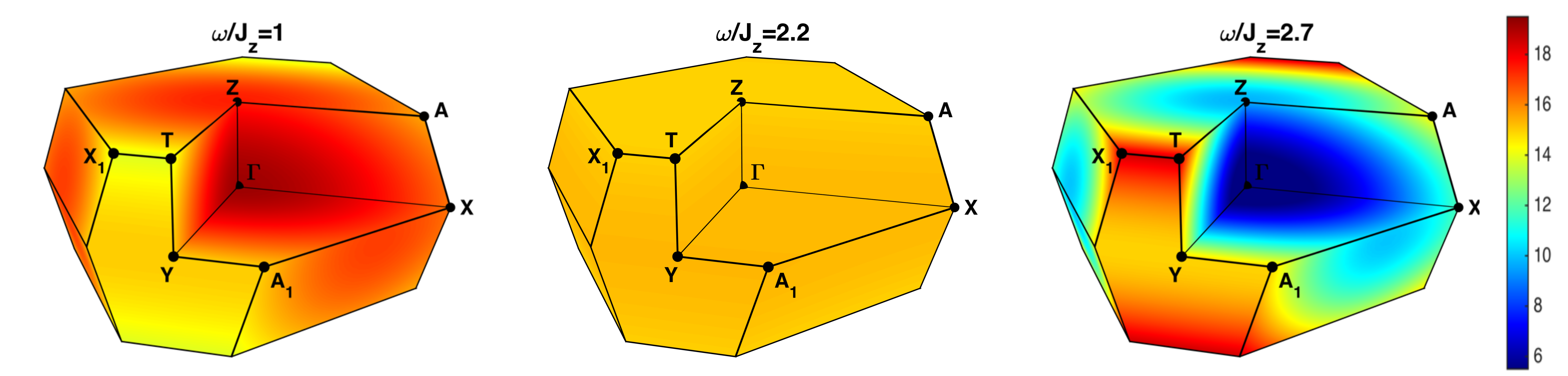}
\caption{Dynamic structure factor for $J_a=1,\ a=x,y,z$ at fixed values of $\omega/J_z$. Intensity is shown for high symmetry planes in the Brillouin zone. The left and right figures show complementary momentum dependence while the center figure shows an `inflexion point' at the transition between the two. Left and right figures are representative of the response below/above the ``inflection point'' at $\omega/J_z \approx 2.2$} \label{fig: 3D plots}
\end{figure*}

\paragraph{\textit{Results.}}
It was suggested recently \cite{Kim} that the effective Hamiltonian for $\beta$-$\mathrm{Li_2 Ir O_3}$ is likely to be in the regime close to the ferromagnetic isotropic point, and so we concentrate on this point in parameter space in the following.
We note that behaviour is different in the opposite limit of one dominant exchange.
For example, the component of response $S^{aa}$ corresponding to the dominant interaction $J_a$ has a sharp $\delta$-function contribution from a pure flux excitation whose presence/absence  gives rise to a {\it dynamical} phase diagram similar to the 2D case \cite{PRL} (not shown).

Fig.~\ref{fig: full struct} shows the $\omega$-dependence of the DFS along cuts in the Brillouin zone (BZ). The response vanishes below the flux-gap, although this is a \textit{gapless} QSL phase. Gross features of DSF are remarkably similar to those of the 2D honeycomb case~\cite{PRL}. This is due to a similar linearly vanishing Majorana DOS both on the 2D honeycomb and 3D hyper-honeycomb lattice, and the fact that the DSF originates from a {\it local} quantum quench despite the extended nature of the flux-loops in 3D. Response above the flux-gap varies smoothly with energy, and falls off rapidly above the fermion bandwidth. Extra fine-structure due to the 3D nature of the lattice is generated from multiple Majorana bands originating from the increased size of the unit cell. 
The momentum dependence of the DSF in the Brillouin zone for three fixed values of $\omega/J_z$ is shown in Fig.~\ref{fig: 3D plots}. 
The limited variation reflects the fact that spin correlations in the QSL are short-range.
We find that there is a striking ``inflexion point'' at energy $\omega/J_z \approx 2.2$ across which the relative intensities in the BZ reverse. At the inflexion point, the response is essentially independent of wavevector across the  BZ. 
The reversal can be traced back to mixing of different contributions from nearest-neighbour (n.n.) and same-site (s.s.) correlators as shown in Fig.~\ref{fig: components}. Note the characteristic peak (dip) for the s.s. (n.n.) components around $\omega=2.7 J_z$, which is related to the corresponding peak in the DOS as shown in the inset. 
Furthermore, at the edge of the BZ there is dependence only on the $X_1 - T$, $T-A_1$ direction which is true for all values of $\omega$. These features are also independent of small anisotropies of the $J_a$.

Finally, the results shown in Fig.~\ref{fig: powder average} compare the powder averaged DSF for the  ferromagnetic and the antiferromagnetic isotropic points. The notable distinction is that the intensity peak shifts from $|\mathbf{q}|=0$ for the former to the boundary of the BZ (around $|\mathbf{q}| \approx 3$) for the latter. Compared to the honeycomb lattice case~\cite{Banerjee2015} we find three, instead of two, broad modes in energy.

\paragraph{\textit{Discussion.}}
The presence of a flux-gap and the broad response at higher energies provide measurable signatures of fractionalized spin-excitations -- emergent gauge fluxes and Majorana fermions -- in INS experiments. While  $\beta$-$\mathrm{Li_2 Ir O_3}$ materials are long-range ordered at low temperatures, it might be nevertheless possible to observe the signatures of fractionalisation in dynamical scattering experiments. This should hold as long as emergent quasiparticles are only weakly confined, so that they represent the natural degrees of freedom to account for the short-time behaviour, which is governed by a dominant Kitaev interaction.

Despite fundamental differences between the Kitaev QSL phases in 2D and 3D  (point like fluxes versus loops, cross-over versus true phase transition at finite temperature) we find that increased dimensionality does not affect the qualitative behaviour of the dynamical structure factor. The latter is surprisingly similar on the 2D honeycomb and 3D hyper-honeycomb lattices, in contrast to other response functions such as Raman scattering \cite{Knolle2014,Perreault2015}, which shows {\it qualitative} differences between 2D and 3D. This observation suggests that INS is the method of choice for  a spectroscopic probe
of Majorana fermions. The way INS couples to these is sufficiently complex to provide a picture of their full spectrum, but simple enough to allow for a direct measure of their properties. 

Indeed, the insensitivity of spin-correlations to dimensionality suggests a great level of universality for Kitaev QSLs on the harmonic honeycomb series. This is related to the fact that on all these polymorphs the dynamical response originates from a local quantum quench, which arises from the hierarchy of the fractionalized quasiparticles - gapless Majorana fermions and static fluxes. 

This viewpoint in terms of emergent degrees of freedom and their interplay in a local quench setting~\cite{BaskaranExact} thus appears as the natural language for probing fractionalisation in Kitaev spin liquids. It immediately suggests that instead of changing dimensionality, qualitative differences will rather result from a change of the low energy DOS. In particular, studying the DSF for a QSL with entire gapless zero energy surfaces \cite{HermannsQSL} or only Weyl points \cite{HermannsWeyl} is an obvious and interesting subject for future research, which can be tackled with the methods we have developed. More ambitiously, it will be of interest to see how magnetic instabilities -- in particular those resulting from integrability breaking interactions -- will dependent on dimensionality on one hand, and the properties of the matter fermions on the other. Such studies will be necessary to make detailed, fully quantitative contact with experiment. However, in the presence of integrability-breaking perturbations, the methodological situation -- in the absence of the exact approach we have developed for the pure Kitaev models -- seems to be considerably more daunting.

\begin{figure}[b]
\includegraphics[width=.5\textwidth]{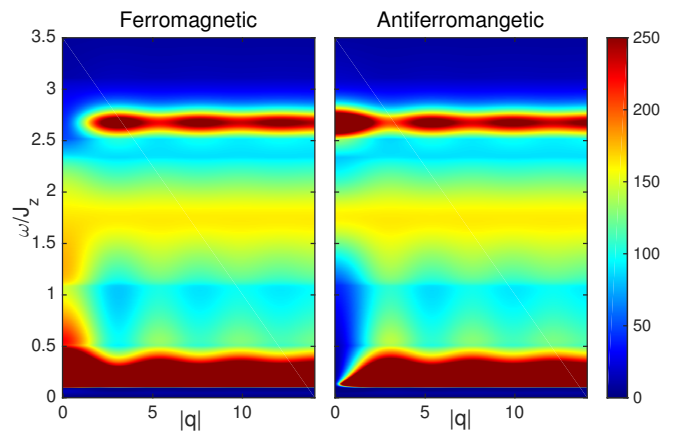}
\caption{Powder averaged dynamical structure factor shown on a linear colour scale (all intensity above 250 (arb. units) in dark red). The results are obtained by integrating angular dependence of $\mathbf{q}$ in $S(\mathbf{q},\omega)$. } \label{fig: powder average}
\end{figure}

\paragraph{\textit{Acknowledgements.}} 
We acknowledge helpful discussions with R.~Coldea, M.~Hermanns, S.~Trebst and S.~Bhattacharjee. The work of J.K.~is supported by a Fellowship within the Postdoc-Program of the German Academic Exchange Service (DAAD). J.T.C. is supported in part by EPSRC Grant No.~EP/I032487/1, D.K.~is supported by EPSRC Grant No.~EP/M007928/1. This collaboration was supported by the Helmholtz Virtual Institute `New States of Matter and their Excitations.'

\bibliography{references}

\end{document}